\documentclass{pasa_rpn}%

\usepackage{graphicx}

\title[Seven Sisters]{Why are there  Seven Sisters?}
\author[Norris \& Norris]{Ray P. Norris$^{1,2}$ \&
Barnaby R. M. Norris$^{3, 4, 5}$\\
\affil{$^1$  Western Sydney University, Locked Bag 1797, Penrith South, NSW 1797, Australia}
\affil{$^2$ CSIRO Astronomy \& Space Science, PO Box 76, Epping, NSW 1710, Australia}
\affil{$^3$ Sydney Institute for Astronomy, School of Physics, Physics Road, University of Sydney, NSW 2006, Australia}
\affil{$^4$ Sydney Astrophotonic Instrumentation Laboratories, Physics Road, University of Sydney, NSW 2006, Australia}
\affil{$^5$ AAO-USyd, School of Physics, University of Sydney, NSW 2006, Australia}}

\usepackage{aas_macros}
\usepackage{hyperref} 
\hypersetup{colorlinks,citecolor=blue,linkcolor=blue,urlcolor=blue}

\hypersetup{draft}

\usepackage{aas_macros}
\usepackage{hyperref} 
\hypersetup{colorlinks,citecolor=blue,linkcolor=blue,urlcolor=blue}

\usepackage{times,epsfig}
\usepackage{graphics,graphicx}
\usepackage{amsmath}
\usepackage{amsfonts}
\usepackage{psfig}

\usepackage{mathrsfs}
\usepackage{amssymb}
\usepackage{bm}
\usepackage{graphicx}

\newcommand{\kms}{\mbox{km\,s$^{-1}$}}
\def\kms {\ifmmode{{\rm ~km~s}^{-1}}\else{~km~s$^{-1}$}\fi}

\def\lsun {\ifmmode{{\rm ~L}_\odot}\else{~L$_\odot$}\fi}

\newbox\grsign \setbox\grsign=\hbox{$>$} \newdimen\grdimen \grdimen=\ht\grsign
\newbox\simlessbox \newbox\simgreatbox
\setbox\simgreatbox=\hbox{\raise.5ex\hbox{$>$}\llap
 {\lower.5ex\hbox{$\sim$}}}\ht1=\grdimen\dp1=0pt
\setbox\simlessbox=\hbox{\raise.5ex\hbox{$<$}\llap
 {\lower.5ex\hbox{$\sim$}}}\ht2=\grdimen\dp2=0pt

\def\lsim{\mathrel{\rlap{\lower4pt\hbox{\hskip1pt$\sim$}}
    \raise1pt\hbox{$<$}}}                
\def\gsim{\mathrel{\rlap{\lower4pt\hbox{\hskip1pt$\sim$}}
    \raise1pt\hbox{$>$}}}                

\def\apj {{\it Ap.~J.}}
\def\apjl {{\it Ap.~J.\ (Letters)}}
\def\apjs {{\it Ap.~J.\ Suppl.}}

\def\aap {{\it Astr.~Ap.}}

\def\araa {{\it Ann.\ Rev.\ Astr.\ Ap.}}

\def\mnras {{\it MNRAS}}

\def\nat {{\it Nature}}
\def\pasa {{\it PASA}}

\def\araa{ARA\&A}             
\def\apj{ApJ}                 
\def\apjl{ApJ}                
\def\apjs{ApJS}               
\def\ao{Appl.~Opt.}           
\def\aap{A\&A}                
\def\jrasc{JRASC}             
\def\mnras{MNRAS}             

\def\qjras{QJRAS}             
\def\nat{Nature}              

\def\grtsim{\mathrel{\hbox{\rlap{\hbox{\lower2pt\hbox{$\sim$}}}\raise2pt\hbox{$>$}}}}
\def\lesssim{\mathrel{\hbox{\rlap{\hbox{\lower2pt\hbox{$\sim$}}}\raise2pt\hbox{$<$}}}}

\def\lsim{\,\lower2truept\hbox{${<\atop\hbox{\raise4truept\hbox{$\sim$}}}$}\,}
\def\gsim{\,\lower2truept\hbox{${> \atop\hbox{\raise4truept\hbox{$\sim$}}}$}\,}
\def\simlt{\mathrel{\rlap{\lower 3pt\hbox{$\sim$}}
        \raise 2.0pt\hbox{$<$}}}
\def\simgt{\mathrel{\rlap{\lower 3pt\hbox{$\sim$}}
        \raise 2.0pt\hbox{$>$}}}

\begin{document}

\maketitle%
\begin{abstract}
There are two puzzles surrounding the Pleiades, or Seven Sisters. First, why are the mythological stories surrounding them, typically involving seven young girls being chased by a man associated with the constellation Orion, so similar in vastly separated cultures, such as the Australian Aboriginal cultures and Greek mythology? Second, why do most cultures call them “Seven Sisters" even though most people with good eyesight see only six stars? Here we show that both these puzzles may be explained by a combination of the great antiquity of the stories combined with the proper motion of the stars, and that these stories may predate the departure of most modern humans out of Africa around 100,000 BC. 


\end{abstract}


\section{Introduction }
\label{intro}

The  Pleiades, or Seven Sisters, is an open stellar cluster of hot, blue, young stars,  which  
were formed about 115 to 125 million years ago \citep{stauffer98, ushomirsky98}, and they are still surrounded by a blue reflection nebula. The cluster is called the ``Seven Sisters" in many cultures, with a  remarkable similarity in the stories surrounding it. 

The importance of the Pleiades in many cultures has been listed by several authors \citep[e.g.][]{allen99,avilin98, burnham78, dempsey09, krupp94, kyselka93, sparavigna08},  from the first written record by the Chinese in 2357 BC through to the present day. In most cultures, the Pleiades are seen as seven young women, or `daughters'  \citep{krupp94}.

 The oldest representation of the Pleiades is thought to be on the Nebra disk, found in Germany, and constructed  around 1600BC \citep{ehser11}, but that representation consists of six stars arranged symmetrically around a seventh, and  is therefore probably symbolic rather than a literal picture of the Pleiades. 

In Greek mythology, the Seven Sisters are named after the Pleiades, who were the daughters of Atlas and Pleione. 
Their father, Atlas, was forced to hold up the sky, and  was therefore unable to protect his daughters. But
to save them from being raped by Orion the hunter, Zeus transformed them into stars. 
Orion was the son of Poseidon, the King of the sea, and a Cretan princess. 
Orion first appears in  ancient Greek calendars  \citep[e.g.][]{planeaux06}, 
but by the late eighth to early seventh centuries BC, he is said to be making unwanted advances on the Pleiades (Hesiod, Works and Days, 618-623). 

Curiously, similar stories about the Pleiades and Orion are told in Aboriginal Australia. For example, most Aboriginal cultures  associate Orion with a hunter, or a young man, or a group of young men, or a male ceremony, and many have stories in which the  men in Orion are trying to chase and rape the girls of the Pleiades \citep[e.g.][]{massola68, mountford39, mountford76}.  The  similarity between the Aboriginal and Greek stories of the Pleiades and Orion includes three specific elements:
both identify the Pleiades as a group of young girls, both
identify Orion as male, and both say that Orion is attempting to have sex with the girls in the Pleiades. 

These strong similarities  suggest a common origin, which appears to predate European contact with Aboriginal Australia.
 
 This comparison is
particularly interesting because there has been almost no cultural contact between the European (i.e. Greek)  and Aboriginal Australian cultures  from about 100,000 BC, when the ancestors of both cultures migrated out of Africa, until  1788 when the British invaded Australia. Nevertheless, there is a remarkable similarity between the stories in both cultures. \citet{norris09} first suggested that one explanation for this similarity  is that the roots of the Seven Sisters story could date back to 100,000 BC, thus providing a common ancestry for this story in all modern human cultures.

This paper examines this ``Out of Africa'' hypothesis.

A related puzzle concerns the number of stars in the Pleiades.
 Although, in principle, ten stars in the Pleiades are sufficiently bright ($m_v < 6$) to be seen with the naked eye, most people with good eyesight, in a dark sky, see only six stars \citep{kyselka93}. This is not a new phenomenon: even in the third century BC, the Greek poet Aratus of Soli gave the names of the Seven Sisters (Halcyone, Merope, Celaeno, Electra, Sterope, Taygete, and  Maia) but then reported  that ``only six are visible to the eyes'' \citep{krupp91}. Thus, while many cultures regard the cluster as having seven stars, they acknowledge that only six are normally visible, and then have a story to explain why the seventh is invisible.

These ``lost Pleiad'' stories are found in European, African, Asian, Indonesian, Native American and Aboriginal Australian cultures \citep{burnham78, gibson17}. 
In Greek mythology,  one of the sisters, Merope, was ashamed of falling in love with a mortal and  therefore faded from sight \citep{sparavigna08}. 
In Australian Aboriginal mythology, one (or occasionally two) of the sisters has died, is hiding, is too young, or has been abducted, so only six (or five) are visible \citep{fuller14a, kyselka93}. 
 \citet{krupp91} gives a story from the Onondaga Iroquois in which one of the stars sang as they ascended to the sky and thus became fainter. In Islam, the seventh star fell to earth and became the Great Mosque \citep{ammarell15}. It is hard to escape the conclusion that once upon a time there really were seven easily visible stars, one of which is no longer visible.
 \citet{hertzog87} describes this phenomenon as ``the combined testimony of numerous societies, spanning continents and millennia, for a seventh easily visible  ...  Pleiad which subsequently dimmed''

%

\section {The Astronomy of the Pleiades}
The Pleiades is one of the nearest open clusters to the Sun,  at a distance of about 135 pc \citep{melis14}, and one of the youngest, with an age of $\sim$ 115 - 125 million years) \citep{ushomirsky98, stauffer98}. The  Pleiades contains stars spanning a wide range of masses, but the brightest visible stars are all B stars. The dynamics of the cluster as a whole are well-studied \citep[e.g.][]{converse10} but subsequent discussion in this paper is limited to the ten stars that are, in principle, visible to the human eye, with $m_{v} < 6$, listed in Table 1.

\begin{table}
\caption{The visible stars of the Pleiades, taken from the Hipparcos catalog \citep{vanleeuwen09} Those corresponding to the Pleiades of Greek mythology are marked with an asterisk. Asterope is a binary with a separation of 2.5 arcmin, so the two stars are indistinguishable to most human eyes. The combined brightness of the two stars is $V_{mag}=5.66$ }
\begin{tiny}
\begin{tabular}{lllllll}

\hline
name	&	RA			&	dec			&  Vmag	&	PM(RA)	&	PM(delta) &	spectral\\
		&	J2000		&	J2000		&		&	mas/yr	&	mas/yr	&	type\\
\hline
Celaeno*	&	03 44 48.20	&	+24 17 22.5	&	5.45	&	20.73	&	-44.00	&	B7IV \\       
Electra*	&	03 44 52.52	&	+24 06 48.4	&	3.72	&	21.55	&	-44.92	&	B6III   \\    
18 Tau	&	03 45 09.73	&	+24 50 21.7	&	5.66	&	19.03	&	-46.64	&	B8V      \\   
Taygeta*	&	03 45 12.48	&	+24 28 02.6	&	4.30	&	19.35	&	-41.63	&	B6V        \\ 
Maia*		&	03 45 49.59	&	+24 22 04.3	&	3.87	&	21.09	&	-45.03	&	B8III       \\
Asterope*	&	03 45 54.46	&	+24 33 16.6	&	5.76	&	19.44	&	-45.36	&	B8V         \\
Merope*	&	03 46 19.56	&	+23 56 54.5	&	4.14	&	21.17	&	-42.67	&	B6IV        \\
Alcyone*	&	03 47 29.06	&	+24 06 18.9	&	2.85	&	19.35	&	-43.11	&	B7III       \\
Atlas		&	03 49 09.73	&	+24 03 12.7	&	3.62	&	17.77	&	-44.7	0	&	B8III       \\
Pleione	&	03 49 11.20	&	+24 08 12.6	&	5.05	&	18.71	&	-46.74	&	B7p         \\
\end{tabular}
\end{tiny}
\end{table}

B stars are often variable \citep[e.g.][]{waelkens85} and the variability of the Pleiades has been well-studied \citep[e.g.][]{white17}. Several have been observed to be variable, and Pleione is known to be an irregular variable, varying by as much as 0.5 magnitude in the last century \citep{burnham78}.
However, such studies are only sensitive to short-term variability (on a timescale of days to years). On a timescale of tens of millions of years, comparable to the lifetime of the star, the star is expected to gradually increase in luminosity because of its expansion as it  moves across the main sequence \citep{langer12}, but the change over a period of 100,000 years is probably still too small to be visible to the human eye. Additional variability may also be caused by motion of the obscuring dust that veils the Pleiades. Because of these unknown factors, we have no information on how the brightness of these stars varies on timescales of thousands of years.

The proper motion of these stars has been measured accurately by Hipparcos \citep{vanleeuwen09}. Gravitational forces from the mass of the cluster, or from tidal fraction, are negligible over human timescales, and so we  can linearly extrapolate their motion back to prehistoric times, as shown in Figure 1. 

\begin{figure}[h]
\begin{center}
\includegraphics[width=8cm]{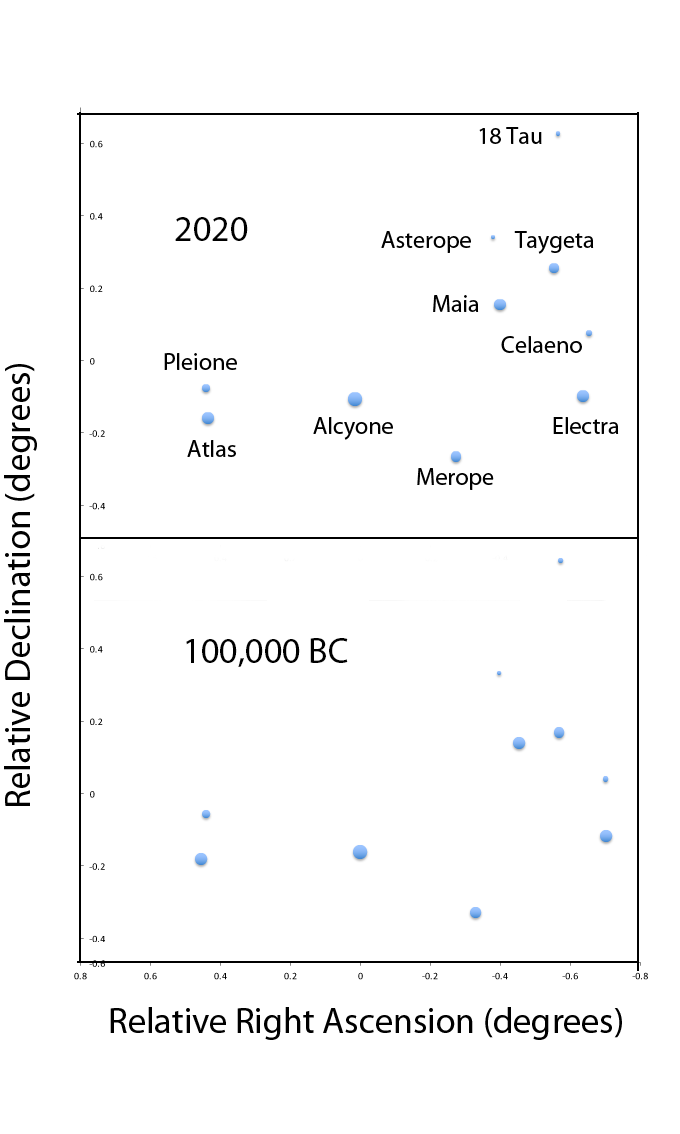} 
\caption{The appearance of the Pleiades at present and at 100,000 BC, assuming linear motion and no variability. The area of each symbol is proportional to ($6 - m_v$), where $m_v$ is the apparent magnitude.}
\label{diagram}
\end{center}
\end{figure}


\section{The Seven Sisters and Orion in Aboriginal Australia}

\subsection{Aboriginal Astronomy}
Astronomy  is a central part of many Aboriginal cultures.
An extensive review, citing all known publications in this field, is given by \cite{norris16}.
 \cite{mountford76}
reported that some Aboriginal people knew the name of every star as faint as fourth magnitude, and knew myths associated with most of those stars. Even now, some elders can name most  stars in the sky visible to the naked eye, and have an intuitive understanding of how the sky rotates over their heads from east to west during the night, and how it shifts over the course of a year  \citep{norris16}. 
 \cite{maegraith32} says that `The most interesting fact about Aboriginal astronomy is that all the adult males of the tribe are fully conversant with all that is known, while no young man of the tribe knows much about the stars until after his initiation is complete ... The old men also instruct the initiated boys in the movements, colour and brightness of the stars.' \cite{dawson81} reported that astronomy  was `considered one of their principal branches of education. ... it is taught by men selected for their intelligence and information'.
 
\subsection{Orion}
\begin{figure}[h]
\begin{center}
\includegraphics[width=6cm]{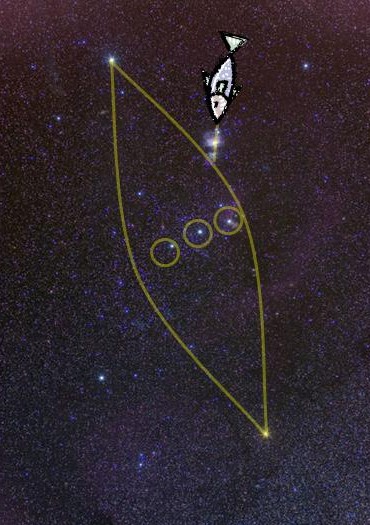} 
\caption{An Australian Aboriginal interpretation of the constellation of Orion, known as ``Djulpan'' in Yolngu, from the Yolngu people of Northern Australia. The three stars of Orion's belt are three young men who went fishing in a canoe, and caught a forbidden king-fish, represented by the Orion Nebula. Drawing by the author based on Yolngu oral and written accounts.}
\label{djulpan}
\end{center}
\end{figure}

Most Aboriginal cultures  associate Orion with a hunter, or a young man, or a group of young men, or a male ceremony, and many have stories in which the  men in Orion are trying to chase and rape the girls of the Pleiades \citep[e.g.][]{massola68, mountford39, mountford76}.  
Examples include:
\begin{itemize}

\item A Yolngu  story that the three stars of Orion's belt are three brothers  in a canoe, with Betelgeuse marking the bow of the canoe, and Rigel the stern. The Orion nebula is a fish, attached by a line to the canoe (Fig.\,\ref{djulpan}). They were blown into the sky by the Sun-woman  as punishment for eating their  totem animal, a king-fish, in violation of  Yolngu law  \citep{davis89, wells73c, norris16}. 

\item The Kaurna story   \citep{gell42, teichelmann40} that Orion is a group of boys who hunt kangaroo and emu on the celestial plain.  

\item The Murrawarri story \citep{mathews94} that Orion wore a belt, carried a shield and stone tomahawk, and their name for the constellation (Jadi Jadi) means either `strong man' or `cyclone'.  

\item The report by \citet{bates25} that people over a great area of central Australia  regarded Orion as a `hunter of women', and specifically of the women in the Pleiades, and that the male initiation ceremony includes an enactment of Orion chasing and raping women. The ceremony may only take place when Orion is {\em not} in the sky, which is consistent with the report  \citep{fuller14a} that, in Kamilaroi culture, Orion's setting in June is associated with the male initiation ceremony.


\end {itemize}

\subsection{Pleiades}
\label{pleiades_section}



The Pleiades are one of the best known features of the Aboriginal sky and its stories have been described extensively \citep[e.g.][]{andrews05, clarke09a, johnson11, norris16} so only a brief description will be attempted here.  In nearly all Australian cultures, the Pleiades are female, and are often   called the Seven Sisters \citep{johnson11, norris16} .  They are generally identified with a group (usually seven, but sometimes six) of young girls, or sisters,  often fleeing from the man or men in Orion (or occasionally from the Moon or another celestial body). 
They are frequently associated with sacred women's ceremonies and stories. The Pleiades are also important as an element of Aboriginal calendars, and in several groups their heliacal rising marks the start of winter.
For example,  \cite{norris16} report an account in which
`Seven sisters come back with turtle, fish, freshwater snakes and also bush foods like yams and berries.The stars come in season when the food and berries come out,  They give Yolngu bush tucker, they multiply the foods in the sea -  that's why Yolngu are happy to see them'.

\citet{johnson11} divides the Pleiades stories  into four groups, to which we add a fifth based on the presence of a protective dingo. The five groups are then as follows:
\begin{itemize}
\item In most areas of mainland Australia,  the Pleiades are portrayed as girls chased by the young men in Orion which is very similar to the  Greek myth
\item In Arnhem Land, stories portray the Pleiades as partners of the men in Orion. 
In some versions of the story from NSW and Victoria, Orion consists of boys who dance at night  to music made by the girls in the  Pleiades  \citep{parker05, smyth78}.
\item In south-west Australia, the stories often feature the girls being protected by their dingoes. Because this  detail is absent in stories from south-east Australia,   \citet{tindale83} argued that the story  predates the arrival of dingoes in Australia in about 5000 BC. The association of the Pleiades with dingoes may also stem from  the harvesting of dingo puppies by several groups as a food source at the heliacal rising of the Pleiades  \citep{harney63, norris07b,  tindale63}.
\item In the Torres Strait Islands, they are  (with Orion) part of the crew of Tagai's boat that perished at sea after Tagai caught them stealing and threw them overboard.
\item In Tasmania, there is no known Pleiades story.
\end{itemize}

Stories in which there are six are usually accompanied by a story explaining how the ``lost Pleiad'' has been raped, or murdered, or has been captured by Orion, or is in hiding from Orion.

 Many  Aboriginal stories refer to the sisters as pursued by the young men in Orion \citep{tindale83}, 
For example,  \cite{harney59} reports a Central Desert version in which
  the girls are being chased towards Uluru by the young Orion men from the North, and escape by fleeing into the sky. 
Similarly, in Kamilaroi culture, Orion is known as the young men who loved, and pursued, the Pleiades
 \citep{mountford76, parker05}.

The Aboriginal stories of Orion and the Seven Sisters are so widespread throughout Australia, and occur in so many different Australian Aboriginal cultures, with local variations, that  these stories are probably thousands of years old, certainly predating  the European occupation of Australia \citep{johnson11}. 
 
 \section{Human Perception of the Pleiades}

Although most people see six stars, some see far more. For example, the first non-Aboriginal Australian astronomer, William Dawes,  claimed to be able to see 13 stars \citep{burnham78}, so evidently was able to see stars fainter than sixth magnitude. There are several other accounts of individuals with exceptional eyesight who can see large numbers of stars. Nevertheless,  most people  see  six stars, a few can see eight, and rarely, those with exceptional eyesight see even larger numbers of stars. However, there is significant disagreement over {\it which} stars are included in the Seven Sisters.

The ``Seven Sisters'' of Greek mythology are unambiguous and are indicated by asterisks in Table 1. 
Most modern people with good eyesight can easily see the brightest five: Alcyone, Merope, Electra, Maia, and Taygeta, all of which are  $m_v =4.3$ or brighter. Atlas ($m_v =3.6$) is often included as one of the Seven Sisters even though, in Greek mythology, Atlas is the father of the sisters.


Pleione is the next brightest star (at $m_v =5.05$), and so is the obvious candidate for the seventh star, although in Greek mythology she is the mother of the Seven Sisters. However, Pleione is very close to the bright star Atlas, making it hard to see, as will be discussed below.
The remaining three stars (Celaeno, 18 Tau, and Asterope are all very faint (at $m_v =5.45$ or fainter),  close to the human limit of sensitivity, and cannot be seeen by most people. Thus the six stars of the Seven Sisters as pointed out by many contemporary observers \citep[e.g.][]{king14} are Alcyone, Merope, Electra, Maia,  Taygeta, and Atlas, with a seventh, Pleione, just visble to those with good eyesight

\section{The Physiology of Seeing Stars}
There are several distinct physiological effects that limit the human perception of stars.

First, the sensitivity of the human eye limits the vision of most people, in a dark sky, to stars brighter than sixth magnitude. Indeed, the system of measuring a stars brightness by its ``magnitude'' was initially based on defining a sixth magnitude star as one that was just visible to the human eye \citep{heifetz04}.

Second, the resolving power of the eye, in bright light, is limited to about one arcmin (which is the distance between the arms of the E on the bottom line of an optometrist's Snellen chart) \citep{yanoff09}, so that two stars closer than this will appear as one. This is a few times worse than might be expected according to the Rayleigh criterion for a diffraction-limited aperture, because of aberrations in the eye.

If these were the only two effects, then most humans would see the ten stars listed in Table 1, as they are all brighter than sixth magnitude and all separated from each other by at least one arcmin. Most people are unable to see Pleione because of two other factors.

First, the one arcmin resolution  is only obtained in bright light, when the resolution, or point spread function (psf), of the human eye is dominated by the cones of the retina. In faint light, human vision relies more on the rods that are sparsely distributed around the retina, and have a much broader psf. 

A second  factor, called ``glare function" by physiologists,  is what prevents you from seeing details next to car headlights pointing at you. The glare function  depends on the dynamic range and psf of the human eye. Imperfections in the human eye give it a psf which has a broad base a few arcmin wide \citep{ginis12}, which in turn limits the dynamic range.  As a result, faint stars cannot be seen within a few arcmin of bright stars. The precise value of the measured  half-width half-maximum (HWHM, at which the psf falls to half of its peak value) of the human psf depends on age, ethnicity, eye colour, and pupil dilation. For example,  Australian Aboriginal people have statistically better acuity than Europeans \citep{taylor81}, although it is not known whether this affects the glare function. Here we assume the results from Fig 5 of \citet{ginis12}, 
from which HWHM appears to be in the range 3 -- 4 arcmin for most people. 

Pleione is five arcmin from the star Atlas, which is about four times brighter than Pleione, and the resulting glare from Atlas prevents most people from seeing  Pleione.

\section{Discussion}

\subsection{The Lost Pleiad}
Although the Pleiades do not appear as seven stars to most humans, could they have  appeared as seven stars in the past? 

There are two potential reasons why they may have done.

First, we have already noted that   many of the Pleiades are B stars, which are often variable. While we have no evidence of any long-term major changes in brightness, and the long-term variability of B stars is poorly understood, we cannot discount the possibility that one of the  faint stars was much brighter in the past. 

\begin{figure}[h]
\begin{center}
\includegraphics[width=6cm]{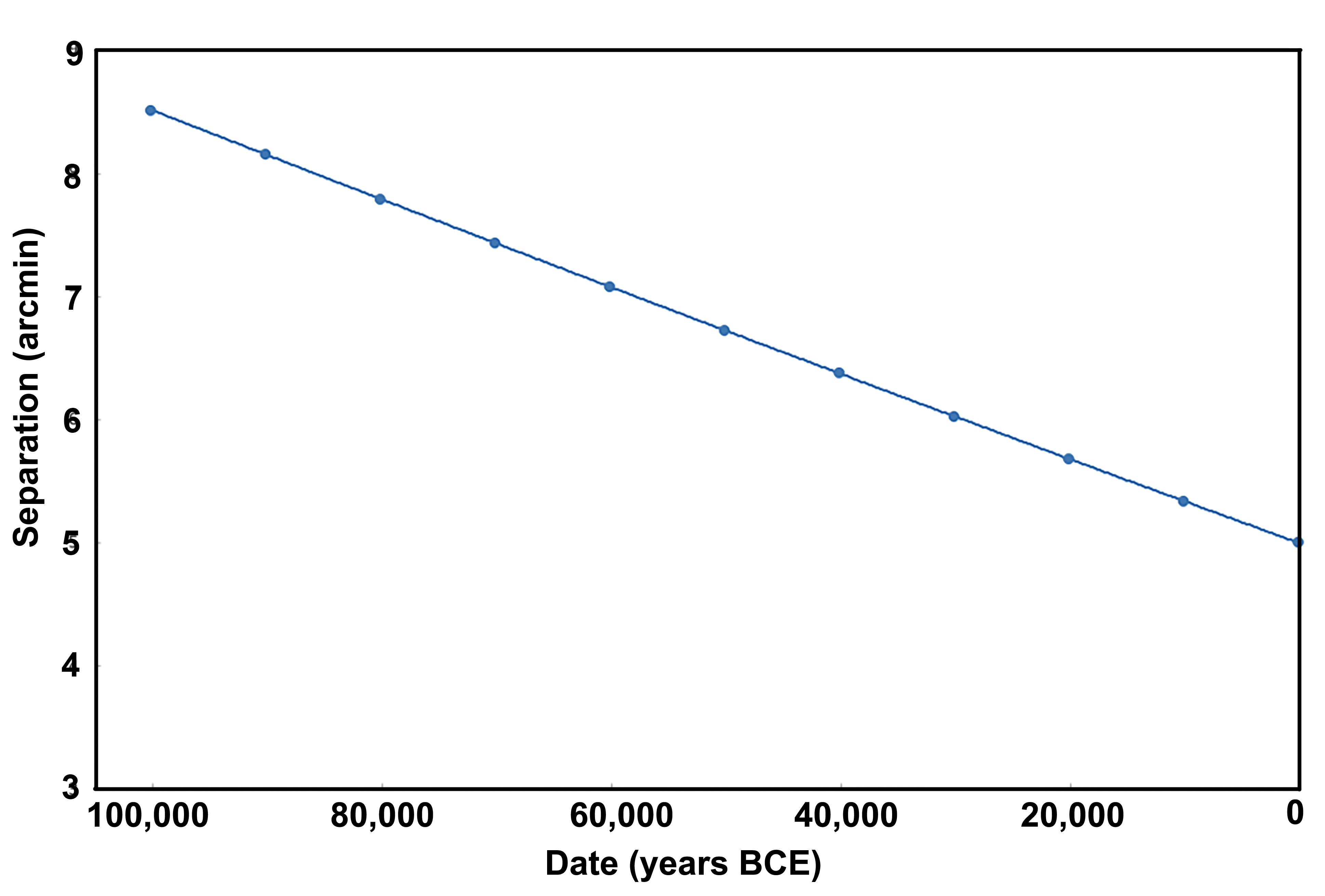} 
\caption{The separation of Atlas and Pleione as a function of time. }
\label{plot}
\end{center}
\end{figure}
\begin{figure}[h]
\begin{center}
\includegraphics[width=6cm]{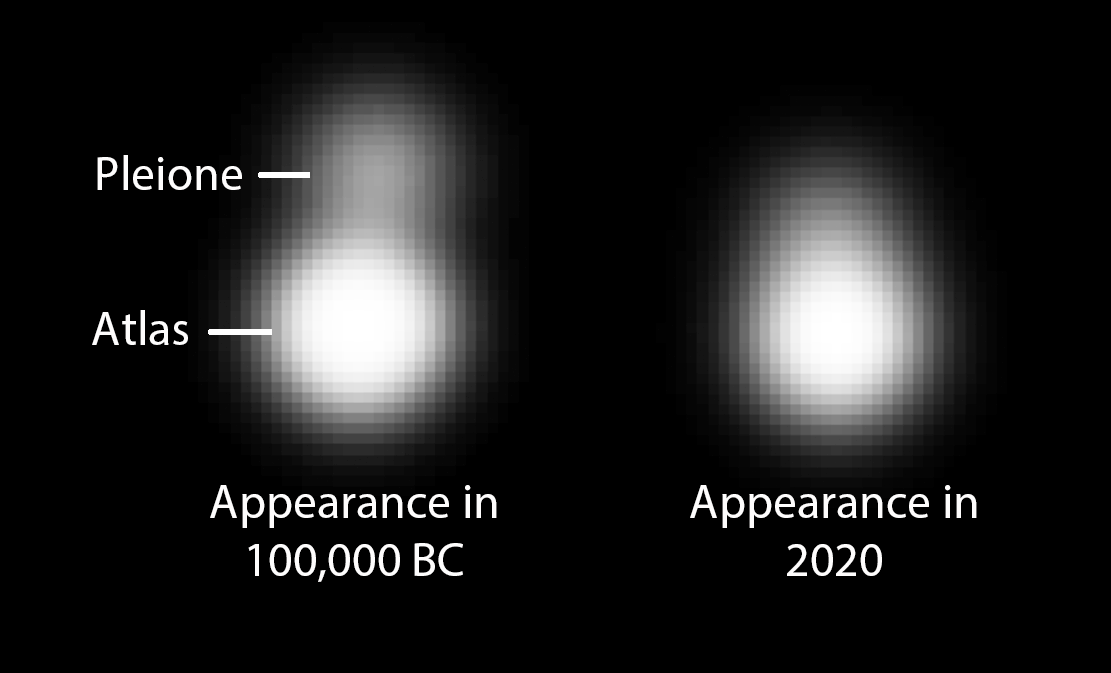} 
\caption{The simulated visual appearance of Atlas and Pleione at the current epoch and at 100,000 BC, as viewed by individuals with a psf HWHM of 3 arcmin.}
\label{appearance}
\end{center}
\end{figure}

Here we suggest an additional reason. Because of Pleione’s measured proper motion,  Pleione was further from Atlas in the past, as shown in Figure \ref{plot}. In 100,000 BC it was 8.4 arcmin away, significantly decreasing the glare from Atlas. Figure \ref{appearance} shows a simulated image of the two stars for an individual with HWHM of 3 arcmin. Even ignoring variability, Pleione was visible as a separate star from Atlas in 100,000 BC, so that the Pleiades would appear as seven stars to normal human eyes 

\subsection{Out of Africa}

The ancestors of Aboriginal Australians left  Africa in about 100,000 BC.  DNA and archaeological studies  \citep{rasmussen,harvati19} show that they were closely related to the ancestors of modern Europeans who left Africa at around the same time.  
The Australians followed the coast of India and China, crossed through Papua New Guinea, 
and arrived in Northern Australia, \citep{hudjashov}, 
probably in a single wave   
at least 40,000 years ago  \citep{oconnell04}.  Radiocarbon dating of Mungo Man showed that they had reached NSW by 40,000 BC  \citep{bowler03}.
A number of recent DNA studies \citep[e.g.][]{nagle17} place the departure date from Africa around 100,000 BC and the arrival date in most of Australia at about 50,000 BC.

From 50,000 BC onwards the Aboriginal people enjoyed a continuous, unbroken culture, with very little contact with outsiders, other than annual visits from Macassan trepang-collectors to the far north of Australia over the last few hundred years.  Aboriginal culture evolved continuously, with no discontinuities or significant outside influences,  until the arrival of the British in 1788, making Aboriginal Australians among the oldest continuous cultures in the world  \citep{mcniven05}.

When the Australians and Europeans  were last together, in 100,000 BC,  the Pleiades would have appeared as seven stars. Given that both cultures refer to them as ``Seven Sisters'', and that their stories about them are so similar, the evidence seems to support the hypothesis that the ``Seven Sisters'' story predates the departure of the Australians and Europeans from Africa in 100,000 BC.

\section{Conclusion}

We have  shown the great similarity between the Aboriginal and Greek stories of the Pleiades and Orion. Specifically, both (in common with many other cultures) predominantly:
\begin{itemize}
\item call the cluster "Seven Sisters", although most humans nowadays see six stars, and then have stories to explain why the seventh is invisible.
\item identify the Pleiades as a group of young girls
\item identify Orion as hunter, or young man, or group of young men
\item have stories in which Orion is attempting to catch or rape the girls in the Pleiades
\end{itemize}
These strong similarities  suggest a common origin, which appears to predate European contact with Aboriginal Australia. This similarity includes an insistence on there being seven stars, even though only six are visible to most people, together with a story to explain the ``lost Pleiad''. The evidence presented above  shows that, because of the proper motion of Pleione, the Pleiades would indeed have appeared as seven stars to most humans in 100,000 BC.
We conclude that the Pleiades/Orion story  dates back to about 100,000 BC, before our ancestors left Africa, and was carried by the people  who left  Africa to become Aboriginal Australians, Europeans, and other nationalities. 

\section{Acknowledgements}
We acknowledge and pay our respects to the traditional owners and elders, both past and present, of all the Indigenous groups mentioned in this paper. We thank Simon O'Toole and Norbert Langer for helpful advice on the variability of the stars of the Pleiades, and Miroslav Filipovic for information about Serbian astronomy. We thank Harilaos Ginis for a helpful discussion about the acuity of the human eye.


\end{document}